\documentclass[reprint,amsmath,amssymb,aps]{revtex4-1}
\usepackage{graphicx,subfigure,amsmath,amsfonts,amssymb,multirow,extarrows,bm}
\usepackage{graphicx}% Include figure files
\usepackage{dcolumn}% Align table columns on decimal point
\usepackage{bm}% bold math
\usepackage[colorlinks,linkcolor=blue,citecolor=blue,urlcolor=blue ]{hyperref}

\usepackage{multirow}
\usepackage{subfigure}
\usepackage{gensymb}
\usepackage{acronym}

\begin{document}
\title{Tetrahedron Constellation of Gravitational Wave Observatory}

\author{Hong-Bo Jin$^{1,2}$}\email[]{Email: hbjin@bao.ac.cn}
\author{Cong-Feng Qiao$^{2,3}$}\email[]{Email: qiaocf@ucas.ac.cn}
\affiliation{\begin{footnotesize}
	    ${}^1$National Astronomical Observatories, Chinese Academy of Sciences, Beijing 100101, China \\
		${}^2$The International Centre for Theoretical Physics Asia-Pacific, University of Chinese Academy of Sciences, Beijing 100190, China \\
		${}^3$The School of Physical Sciences, University of Chinese Academy of Sciences, Beijing 100049, China
\end{footnotesize}}

\date{\today}

\begin{abstract}
For the first time, we have introduced the Tetrahedron Constellation of Gravitational Wave Observatory (TEGO) composed of four identical spacecrafts (S/Cs).
The laser telescopes and their pointing structures are mounted on the S/C platform and are evenly distributed at three locations 120 degrees apart. 
These structures form automatically a stable mass center for the platform. 
The time delay interferometry (TDI) are used to suppress the frequency noise of  Gravitational Wave (GW) detector. The unequal arm Michelson TDI configuration and the Sagnac TDI configuration are equally effective at eliminating the laser frequency noise based on the TEGO configuration. 
Furthermore, comparing to the configurations of LISA, Taiji, and TianQin, the TEGO has more combinations of optical paths in its TDI system sensitive to GW signals.
The six arms of TEGO are simultaneously sensitive to the six polarization modes of GWs. The sensitivity implies that GW modes beyond the predictions of general relativity (GR) can be detected directly. For instance, a scalar longitudinal mode of GWs, which is not predicted by GR, has been identified as a dominant polarization component. This mode is found to be evident in the response amplitudes of the TEGO arms, such as between S/C1 and S/C4, and S/C3 and S/C4, at certain orbital positions.  
\end{abstract}
\maketitle

\section{Introduction}
In general relativity (GR), the gravitational wave (GW) polarization modes are tensor named as plus and cross modes, which have been detected from the LIGO\citep{2017NatCo...814906S}, Virgo\citep{Virgo:2019juy}, and KAGRA\citep{KAGRA:2018plz} collaborations. The other GW modes beyond GR, called as two scalar polarization modes and two vector modes, have being probed especially with an expected network of Advanced LIGO, Advanced Virgo, and KAGRA\citep{Hagihara:2018azu,Hagihara:2019ihn}. The ability of Taiji\citep{TaijiScientific:2021qgx, Gong:2021gvw}, which is a space-based observatory of GWs, to detect the polarization modes is explored in the parametrized post-Einsteinian framework and it is found that Taiji can measure the dipole and quadruple emission of GWs\citep{Liu:2020mab}. The performance of the LISA-TianQin network for detecting polarizations of stochastic GW background is also investigated\citep{Hu:2024toa}.

In the six polarizations of GWs in a general metric theory of gravitation in four dimensions, the additional polarizations mean that the theory of gravitation should be extended beyond GR, and excludes some theoretical models, depending on which polarization modes are detected\citep{Nishizawa:2009bf}. Thus, the observation of the GW polarizations is used to probe the extended law of gravity and extra dimensions. The detector of GWs responses to the six polarizations of GWs in four-dimensional space, which favors the tridimensional detector of GWs. That is attained from the recent studies of the network of some observatories, such as LIGO, Virgo, and KAGRA or LISA, Taiji and TianQin, etc. 

\begin{figure*}
\includegraphics[width=0.44\textwidth]{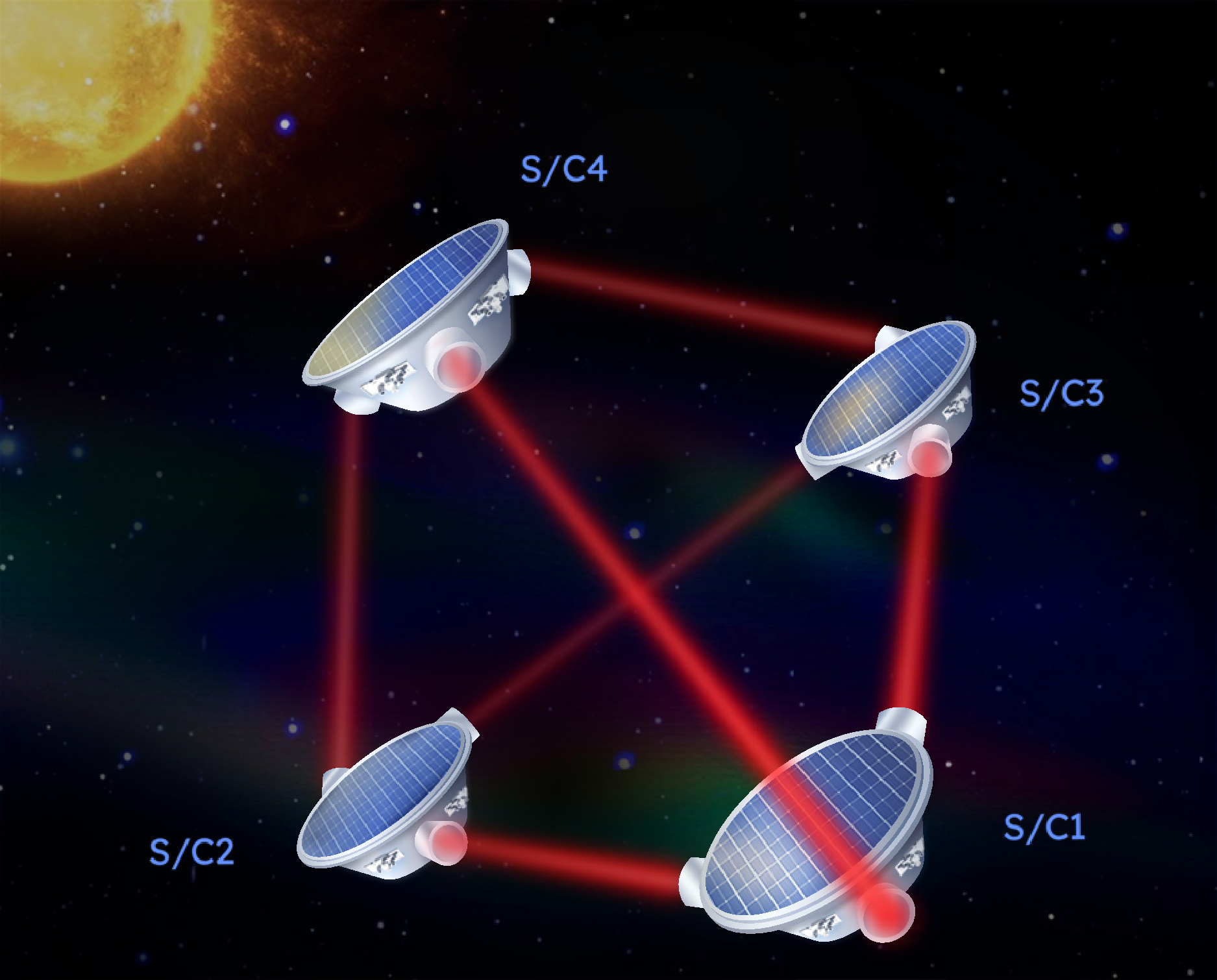}\includegraphics[width=0.54\textwidth]{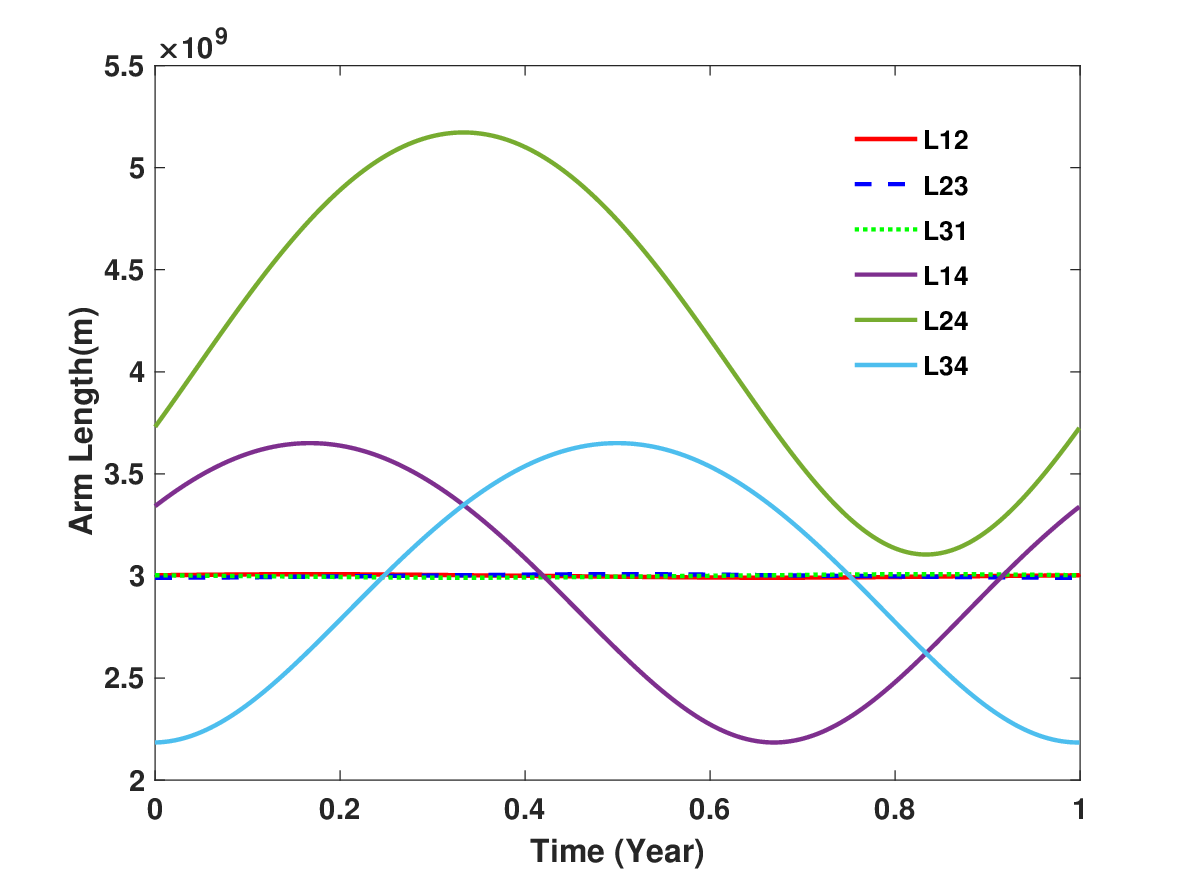}
\caption{(left) The schematic of the tetrahedron constellation of gravitational wave observatory(TEGO). There are four identical S/Cs. The orientation of the laser beam is adjusted by the telescope pointing structure, which is integrated with the telescope and evenly distributed at three locations 120 degrees apart in the spacecraft platform.
(right) Variations of the arm lengths ( Laser path distance between S/Cs ) in 1 year. In the legend, Lmn is between the S/Cm and S/Cn. When mn is 12, it is L12.}
 \label{fig:tdi4}
\end{figure*}
A three-dimensional octahedron constellation, composed of six spacecrafts (S/Cs), has been studied\citep{Wang:2013lna}. Clearly, the 12 arms between these S/Cs are highly redundant for detecting the six polarizations of GWs.  
In this paper, we propose the tetrahedron constellation of gravitational wave observatory (TEGO) with a four triangles composed of four S/Cs for the three-dimensional configuration different from those networks. 
The orientation of the laser beam is adjusted by the telescope pointing structure, which is integrated with the telescope and evenly distributed at three locations 120 degrees apart in the spacecraft platform. 
Comparing to the three S/Cs configuration, the addition of S/C and the mounted telescope are also seen as redundant backups, especially for a telescope, which automatically form a stable mass center of platform with the other two telescopes without the trim mass. This configuration of TEGO is shown in the left of Figure \ref{fig:tdi4}.
The arm lengths of triangles composed of three S/Cs: S/C1 - S/C3 have almost same values. The other arm lengths of triangles composed of three S/Cs with S/C4 and S/C1 - S/C3 have the variable values over a one-year orbital period. That is derived from that the  S/C4 has a longer radial distance than the other S/Cs. The orbit details of S/Cs are found in the Method section. In the right of Figure \ref{fig:tdi4}, the arm lengths between SCs are drawn over a one-year orbital period. TEGO is composed of the triangular S/Cs, whose configuration is LISA-like and senses the displacement and laser noise. That needs the updated data analysis techniques to cancel the those noises. 

In the space-based observatories with a triangle of three spacecrafts, such as LISA\citep{LISA:2017pwj,2022NatAs...6.1334B}, Taiji \citep{Hu:2017mde,Ruan:2020smc, TaijiScientific:2021qgx,Gong:2021gvw} and TianQin\citep{TianQin:2015yph,TianQin:2020hid,Gong:2021gvw}, the laser frequency noise are dominant. The heterodyne interferometry with unequal arm lengths and independent phase-difference readouts has been proposed to cancel the frequency noise\citep{Dhurandhar:2011ik,Tinto:2014lxa}. The properly time-shifting and linearly combining independent Doppler measurements are applied, which has been called time-delay interferometry (TDI)\citep{Dhurandhar:2011ik,Tinto:2014lxa}. 
In the paper\citep{Wang:2017aqq}, the optical path differences of the second-generation TDI were calculated for LISA, TAIJI, and LISA-like missions. These differences are well below the respective limits, below which the laser frequency noise must be suppressed. In the space-based observatories with a triangle of three spacecrafts, such as LISA\citep{LISA:2017pwj,2022NatAs...6.1334B}, Taiji \citep{Hu:2017mde,Ruan:2020smc, TaijiScientific:2021qgx,Gong:2021gvw} and TianQin\citep{TianQin:2015yph,TianQin:2020hid,Gong:2021gvw}, the laser frequency noise are dominant. The heterodyne interferometry with unequal arm lengths and independent phase-difference readouts has been proposed to cancel the frequency noise\citep{Dhurandhar:2011ik,Tinto:2014lxa}. The properly time-shifting and linearly combining independent Doppler measurements are applied, which has been called time-delay interferometry (TDI)\citep{Vallisneri:2005ji,Dhurandhar:2011ik,Tinto:2014lxa}. 
%In the paper\citep{Wang:2017aqq}, the optical path differences of the second-generation TDI were calculated for LISA, TAIJI, and LISA-like missions, which are well below their respective limits which the laser frequency noise is required to be suppressed. 
In this paper, the arm lengths of triangles composed of three S/Cs with S/C4 and S/C1 - S/C3 have the remarkable variable values over a one-year orbital period, that implies the the frequency noise of those optical paths is stronger than the other optical path. Based on the TEGO configuration, The calculation of optical path differences indicate that the unequal arm Michelson TDI configuration and the Sagnac TDI configuration are equally effective at eliminating the stronger laser frequency noise.

Given the periodic motion of GW detectors along their solar orbits, the amplitude of a GW source at a particular sky position is modulated according to the detector's orbital position over a one-year period. Clearly, the polarization directions of GWs relative to the detector also change with the detector's position in orbit. Furthermore, when a detector reaches a specific orbital position, some GW polarizations may remain undetected if the polarization direction of the GWs is perpendicular to an arm of the GW detector. Since the three-dimensional TEGO configuration possesses six arms, it is always possible to find an arm that is not perpendicular to a given polarization direction of GWs. Consequently, the six polarization modes of GWs can be sensitively detected by the six arms of TEGO at any orbital position.
From the response amplitudes of the TEGO arms in a one-year period, a scalar longitudinal mode of GWs has been identified as a dominant polarization component at certain orbital positions.  
\section{methods}
\subsection{Detector Orbit} \label{Sdr_oe}%---------------------

Referring to LISA orbit, S/Cs of TEGO orbit Kepler\citep{Rubbo:2003ap}, the arm length is L=3$\times 10^9$ m between S/C1, S/C2 and S/C3. Each spacecraft (S/C) positions are expressed as a function of time. To second order in the eccentricity, the Cartesian coordinates of the spacecraft are given by
\begin{eqnarray} \label{keporb}
    x(t) &=& R \cos(\alpha) + \frac{1}{2} e R\Big( \cos(2\alpha-\gamma) - 3\cos(\gamma) \Big) \nonumber\\
    && + \frac{1}{8} e^2 R\Big( 3\cos(3\alpha-2\gamma) - 10\cos(\alpha) - 5\cos(\alpha-2\gamma) \Big) \nonumber\\
    y(t) &=& R \sin(\alpha) + \frac{1}{2} e R\Big( \sin(2\alpha-\gamma) - 
    3\sin(\gamma) \Big) \nonumber\\
    && + \frac{1}{8} e^2 R\Big( 3\sin(3\alpha-2\gamma) - 10\sin(\alpha) + 5\sin(\alpha-2\gamma) \Big) \nonumber\\
    z(t) &=& -\sqrt{3} e R\cos(\alpha-\gamma)  + \sqrt{3} e^2 R \Big( \cos^2(\alpha-\gamma)\nonumber\\
   && + 2\sin^2(\alpha-\gamma) \Big) 
\end{eqnarray}
where $R = 1$ AU is the radial distance (the distance from the sun) to the center of the triangle from S/C1 to S/C3.  For S/C4 above the triangle $R$ is (1 + 0.0163738) AU. 0.0163738 is best-fit value when the mean value of arm length is near L=3 $\times 10^9$ m in 1 year. $\alpha = 2\pi t/year + \kappa$ is the orbital phase of the guiding center, and $\gamma = 2\pi m/3 + \Lambda$ ($m=0.0,1.0,2.0,2.5$) is the relative phase from S/C1 to S/C4.  The parameters $\kappa$ and $\Lambda=0$ give the initial ecliptic longitude and orientation of constellation. The orbital eccentricity $e$ is computed based on the arm length: $e = L/(2 \sqrt{3} R )$.
By setting the mean arm-length equal to those of the TAIJI baseline, $L = 3 \times 10^9$ m, the spacecraft orbits are found to have an eccentricity of
$e=0.005789$, which indicates that the second order effects are down by a factor of 100 relative to leading order.

The fourth S/C follows an orbit with the same parameters as the other three, with the only difference being a slightly higher radial distance for the fourth S/C compared to the others. Consequently, its orbit is also stable. As an effect, the distances between the fourth S/C and the other three vary obviously for the one-year orbital period.

\subsection{Time Delay Interferometry}\label{TDI}
TDI combinations, used in this paper, are unequal- arm Michelson and Sagnac, which are shown in FIG. 4 of the paper\citep{Vallisneri:2005ji}.
%, which are used to cancel the frequency noise in a stationary unequal-arm interferometry\citep{Vallisneri:2005ji,Dhurandhar:2011ik,Tinto:2014lxa}.
Specially, for a moving interferometer,  the second-generation TDI combinations are used to further cancel the frequency noise\citep{Vallisneri:2005ji,Dhurandhar:2011ik,Tinto:2014lxa,Wang:2017aqq}. 
As an example, as the distances between the fourth S/C and the other three have remarkable variable than the distances between the triangular S/Cs, the relevant Laser frequency noise are more remarkable.
The cancelation of Laser frequency noise can be estimated by the difference in optical path length between the two paths as follows.

There are two split laser beams to go to Path 1 and Path 2, which interfere at their end path. 
For the unequal-arm Michelson TDI configuration, one laser beam originates from S/C1, directed towards and received by S/C2, where it optically phase-locks with a local laser; the phase-locked laser beam is then directed back to S/C1, optically phase-locking another local laser there. This process continues along Path 1, which returns to S/C1 via the following route.  And the second laser beam also originates from S/C1 but follows Path 2 before returning to S/C1 to interfere coherently with the first beam\citep{Wang:2017aqq}.
\begin{itemize}
\item Path 1: S/C1$\rightarrow$ S/C2$ \rightarrow$ S/C1 $\rightarrow$ S/C3 $\rightarrow$ S/C1, 
\item Path 2: S/C1$\rightarrow$ S/C3$ \rightarrow$ S/C1 $\rightarrow$ S/C2 $\rightarrow$ S/C1. 
\end{itemize}
In Figure \ref {fig:Michelson}, L12 - L13 means the difference between Path 1 and Path 2.  And the others are similar to L12 - L13, but one laser beam originates from S/C2, S/C3 or S/C4.
In the second-generation Sagnac TDI configuration, the two optical paths are configured as follows:
\begin{itemize}
\item Path 1:  S/C1 $\rightarrow$ S/C2 $\rightarrow$ S/C3 $\rightarrow$ S/C1 $\rightarrow$ S/C3 $\rightarrow$ S/C2 $\rightarrow$ S/C1,
\item Path 2: S/C1 $\rightarrow$ S/C3 $\rightarrow$ S/C2 $\rightarrow$ S/C1 $\rightarrow$ S/C2 $\rightarrow$ S/C3 $\rightarrow$ S/C1.
\end{itemize}
In Figure \ref {fig:Sagnac}, L123 - L132 means the difference between Path 1 and Path 2 above.  And the others are similar to L12 - L13, but one laser beam originates from S/C2, S/C3 or S/C4 in one triangular S/Cs.
In Figure \ref {fig:SagnacMix},  L124 - L134 represents the difference in optical path lengths between the two triangular configurations of S/Cs: S/C1, S/C2, and S/C4; and S/C1, S/C3, and S/C4, which share a common arm, L14.  And the others are similar to L124 - L134, but one laser beam originates from S/C2, S/C3 or S/C4 in two triangular S/Cs.
If the two paths have exactly the same optical path length, the laser frequency noises cancel out. If the difference in optical path length between the two paths is small, the laser frequency noises are largely canceled out\citep{Wang:2017aqq}.

\subsection{GW Signal Model of white dwarf J0806}\label{wavformModel}%---------------------
%=== waveform of monochromatic source ===
The GW signals from white dwarf binary are described by a set of eight parameters: frequency $f$, frequency derivative  $\dot{f}$, amplitude ${\cal A}$, sky position in ecliptic coordinates $(\lambda,\beta)$, orbital inclination $\iota$, polarisation angle $\psi$, and initial orbital phase $\phi_0$ \citep{Cutler:1997ta,roe20,kar21}. The recipes for generating ${\cal A}$, $f$, $\dot{f}$, and $(\lambda,\beta)$ are based on the currently available observations. 
The gravitational wave emitted by a monochromatic source is calculated using the quadrupole approximation \citep{Landau:1962,Peters:1963}.
In this approximation, the GW signals are described as a combination of the two polarizations ($+, \times$)\citep{Korol:2021pun}: 
\begin{eqnarray}\label{eq:hplus_hcross}
    h_+(t) = \mathcal{A}(1+\cos\iota^2)\cos( \Phi(t)) \nonumber\\
    h_{\times}(t) = 2\mathcal{A}\cos\iota\sin(\Phi(t)).
\end{eqnarray}
In above expression,
\begin{eqnarray}
	\mathcal{A} =\frac{2(G\mathcal{M})^{5/3}}{c^{4}D_L}(\pi f)^{2/3}\nonumber\\
    \Phi(t)=2\pi ft + \pi\dot{f}t^2 + \phi_0  \nonumber\\
    \dot{f}=\frac{96}{5}  \pi^{8/3} \left( \frac{G{\cal M}}{c^3} \right)^{5/3} f^{11/3}. 
\end{eqnarray}
where $\mathcal{M}\equiv (m_1 m_2)^{3/5}/(m_1+m_2)^{1/5}$ is the chirp mass, and $D_L$ is luminosity distance, and $G$ and $c$ are the gravitational constant and the speed of light, respectively.

A verification white dwarf binaries signals: J0806 as the identified sources are selected, whose parameters are detailed in the paper\citep{Kupfer:2018jee}. The polarization angle is set $\psi=0$, and initial orbital phase is set $\phi_0=0$. The sky location parameters \mbox{$\lambda [rad], \beta [rad]$ and $ D_L [kpc]$} are 2.1021, -0.0821 and 5 respectively\citep{Kupfer:2018jee}. Expect for two polarizations ($+, \times$), the other four components of GW polarizations cannot calculated with quadrupole approximation \citep{Landau:1962,Peters:1963}. In this paper, the GW waveforms relevant to the other four components ($U,V,B,L$)  of GW polarizations are set to be equal to ($\times$) component, artificially.

\subsection{Polarization Modes}\label{polarizationModel}
In four dimension space-time, GWs have six polarization modes: two transverse-traceless modes [plus ($+$) and cross ($\times$)],
two vector longitudinal modes ($U$ and $V$), a scalar transverse breathing mode ($B$), and a scalar longitudinal mode ($L$).
The total tensor with all six polarization modes is written as
\begin{eqnarray}
\label{HSUM}
\mathbf{H}(t) = h_{+}(t) \boldsymbol{\epsilon}^{+}+h_{\times}(t) \boldsymbol{\epsilon}^{\times}+h_{U}(t)\boldsymbol{\epsilon}^{U} \nonumber\\
 +h_{V}(t) \boldsymbol{\epsilon}^{V}+h_{B}(t) \boldsymbol{\epsilon}^{B}+h_{L}(t) \boldsymbol{\epsilon}^{L} ,
\end{eqnarray}
where $\boldsymbol{\epsilon}^{A}$ ($A=+,\times,U,V,B,L$) are the polarization tensors and $h_{A}$ are the waveforms of the polarization modes.
 
Solar System Barycenter ($SSB$) frame, based on the ecliptic plane, is selected. In $SSB$ frame, the standard spherical coordinates $\left(\theta,\phi\right)$ and the associated spherical orthonormal basis vectors $\left(\bf{e}r,\bf{e}\theta,\bf{e}\phi\right)$ are confirmed. The position of the
source in the sky is parametrized by the ecliptic latitude $\beta =\pi/2 - \theta$ and the ecliptic 
longitude $\lambda = \phi$. The GW propagation vector $\hat{k}$ in spherical coordinates is expressed as
\begin{equation}
    \hat{k} = -\bf{e} r = -\cos\beta\cos\lambda\,\hat{x} - \cos\beta\sin\lambda\,\hat{y} -
    \sin\beta\,\hat{z} \,.
\end{equation}
The reference polarization vectors is the following:
\begin{eqnarray}
    \hat{u} &=& -\bf{e}\theta = -\sin\beta\cos\lambda\,\hat{x} -\sin\beta\sin\lambda\,\hat{y} + \cos\beta\,\hat{z} \nonumber\\
    \hat{v} &=& -\bf{e}\phi = \sin\lambda\,\hat{x} - \cos\lambda\,\hat{y} .
\end{eqnarray}

The polarization tensors are given
\begin{eqnarray}
\label{epsilon}
\boldsymbol{\epsilon}^{+} &=& \mathbf{e}^{+} \cos 2\psi -\mathbf{e}^{ \times}\sin 2\psi  , \nonumber \\
\boldsymbol{\epsilon}^{\times} &=& \mathbf{e}^{+}\sin 2\psi+\mathbf{e}^{ \times}\cos 2\psi  , \nonumber \\
\boldsymbol{\epsilon}^{U} &=& \mathbf{e}^{U}\cos \psi-\mathbf{e}^{V}\sin \psi  , \nonumber \\
\boldsymbol{\epsilon}^{V} &=& \mathbf{e}^{U}\sin \psi +\mathbf{e}^{V}\cos \psi  , \nonumber \\
\boldsymbol{\epsilon}^{B} &=& \mathbf{e}^{B} , \nonumber \\
\boldsymbol{\epsilon}^{L} &=& \mathbf{e}^{L} ,
\end{eqnarray}
where $\psi$ is the polarization angle and the six basis tensors are~\citep{Nishizawa:2009bf}
\begin{eqnarray}
\mathbf{e}^{+} &=& \hat{u} \otimes \hat{u}-\hat{v} \otimes \hat{v} , \nonumber \\
\mathbf{e}^{\times} &=& \hat{u} \otimes \hat{v}+\hat{v} \otimes \hat{u} , \nonumber \\
\mathbf{e}^{U} &=& \hat{u} \otimes \hat{k}+\hat{k} \otimes \hat{u} , \nonumber \\
\mathbf{e}^{V} &=& \hat{v} \otimes \hat{k}+\hat{k} \otimes \hat{v} , \nonumber \\
\mathbf{e}^{B} &=& \hat{u} \otimes \hat{u}+\hat{v} \otimes \hat{v} , \nonumber \\
\mathbf{e}^{L} &=& \hat{k} \otimes \hat{k} .
\end{eqnarray}

\subsection{Detector Response}
The laser link signal is emitted from the  S/C $i$ at $t_i$ moment and received by the  S/C $j$ at $t_j$ moment. 
Detector arm L${}_{i j}$ is the laser link from S/C $i$ to  S/C $j$, whose displacement is $\ell_{i j}(t)$ at t moment.
The gravitational wave travels in the $ \hat k$ direction and ${\mathbf x (t)}$ is the position on the laser link at time t.
By the Eq.~(\ref{deltaL2}), the $\delta \ell_{i j}(t)$ is calculated in a numerical way in the time domain\cite{Guo:2023lzb}.
\begin{equation} \label{deltaL2}
    \delta \ell_{i j}(t) =\frac{1}{2} \hat{r}_{i j}(t) \otimes \hat{r}_{i j}(t) : \int_{t_i}^{t_j} {\bf h}(t-\hat{k} \cdot {\mathbf x (t)}) d t
\end{equation}
where,
$\hat{r}_{i j}(t)$ denotes the unit vector
\begin{equation}\label{rij}
    \hat{r}_{i j}(t) = \frac{{\bf x}_j(t_j) - {\bf
        x}_i(t_i)}{\ell_{i j}(t)} \, .
\end{equation}
and ${\bf h}(t-\hat{k} \cdot {\mathbf x (t)})$ is the GW tensor in the transverse-traceless gauge, whose components are $\boldsymbol{h}_{A}$ ($A=+,\times,U,V,B,L$) from the Eq.~(\ref{HSUM}). 
%The colon here denotes a double contraction, ${\bf a}:{\bf b} = a^{i j}b_{i j}$.
The amplitude of GW, responded to detector arm L${}_{i j}$, is in the following equation,
\begin{equation}\label{Lij}
     \mathbf{H}_{ij}(t) = \frac{\delta \ell_{i j,A}(t)\epsilon^{A}}{\ell_{i j}(t)} ,
\end{equation}
where $\boldsymbol{\epsilon}^{A}$ ($A=+,\times,U,V,B,L$) are the polarization tensors from the Eq.~(\ref{epsilon}). The upper and lower indicator A of the two quantities: ${\delta \ell_{i j,A}}$ and ${\epsilon^{A}}$ are summed.
In the Fig. \ref{fig:polarizations}, $h_{sum}$ for L${}_{i j}$ is $\mathbf{H}_{ij}(t)$ and
 $h_{A}$ for L${}_{i j}$ is a component value equal to ${\delta \ell_{i j,A}(t)\epsilon^{A}}/{\ell_{i j}(t)}$ ( the upper and lower indicator A is only used as a component and not summed).

\begin{figure*}[!htbp]
	\centering 
    \includegraphics[width=0.45\textwidth]{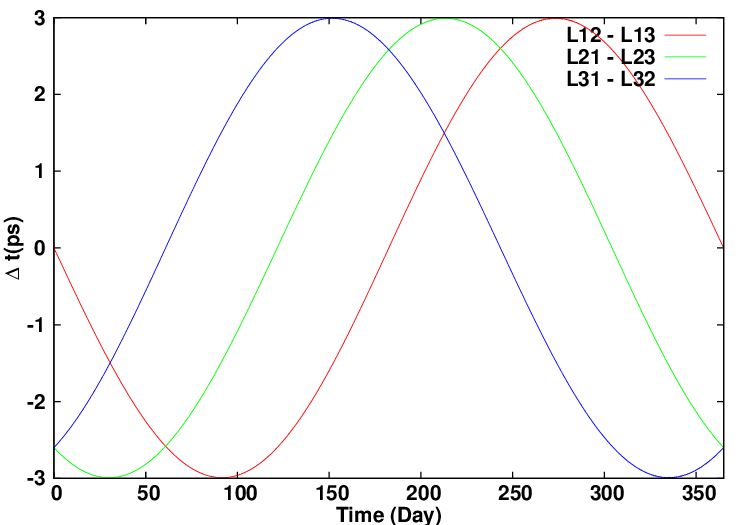}\includegraphics[width=0.45\textwidth]{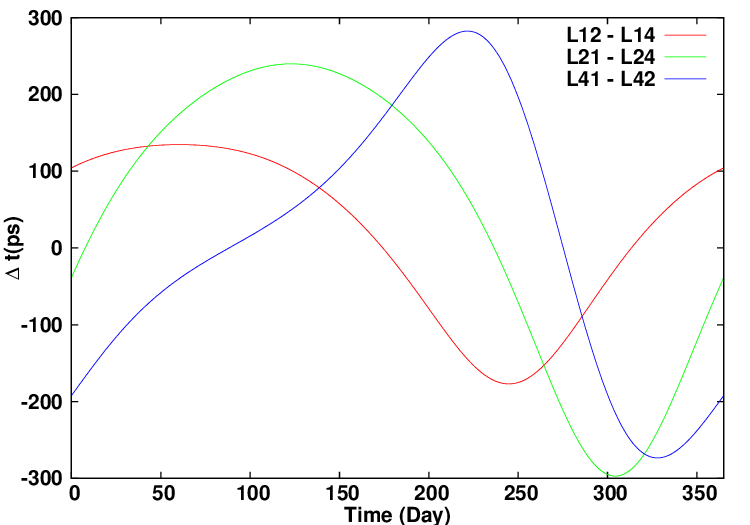} 
     \includegraphics[width=0.45\textwidth]{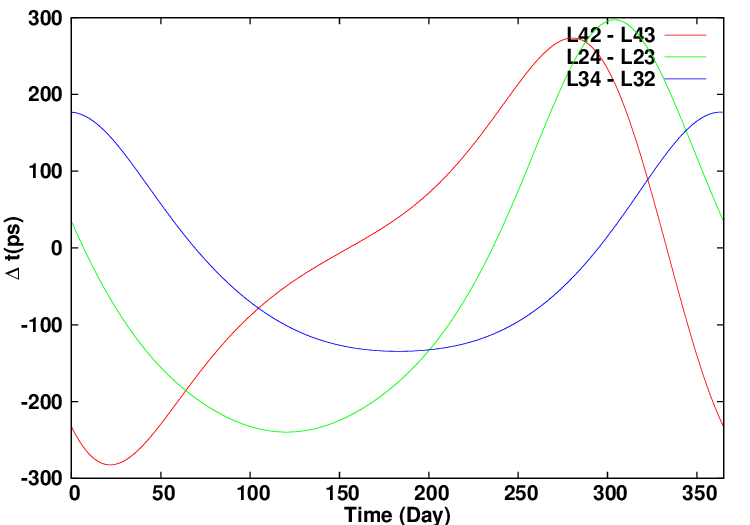}\includegraphics[width=0.45\textwidth]{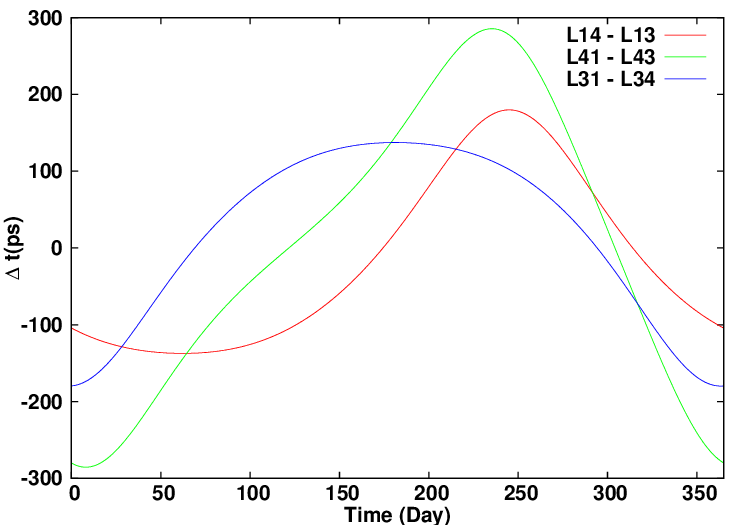}
    \caption{The optical path differences of the second-generation Michelson TDI over a one-year orbital period. The unit of the y-coordinate is the picosecond (ps) and the optical path difference is C$\Delta t$. C is light speed. As an example, in the legend, L12 - L13 means the optical path differences between S/C1$\rightarrow$ S/C2$ \rightarrow$ S/C1 $\rightarrow$ S/C3 $\rightarrow$ S/C1 and S/C1$\rightarrow$ S/C3$ \rightarrow$ S/C1 $\rightarrow$ S/C2 $\rightarrow$ S/C1.}
    \label{fig:Michelson}
\end{figure*}
\section{Results}
Comparing to those triangular detectors, TEGO has more than four times combinations of optical paths of time delay interferometer(TDI) to suppress the frequency noise. 
In this paper, based on the unequal arm Michelson TDI configuration and the Sagnac TDI configuration for three S/Cs, the optical path differences of the second-generation TDI are calculated to check the capability of TDI for the apparently variable arm lengths.
As is seen in Figure \ref{fig:Michelson}, Michelson TDI configuration is from the optical path combination of two arms and the optical path differences are below the limits of noise canceling, compared to the results in the paper\citep{Wang:2017aqq}. In the optical path differences of all the combination, the values relevant to the combination between S/C1, S/C2 and S/C3 are 1\% less than the others, which is seen in the Figure \ref{fig:Michelson}. This result is due to the fact that,  over a one-year orbital period, three arm lengths relevant to S/C4  are changed apparently, that is seen in the right of Figure \ref{fig:tdi4}. 

\begin{figure*}[!htbp]
	\centering 
    \includegraphics[width=0.45\textwidth]{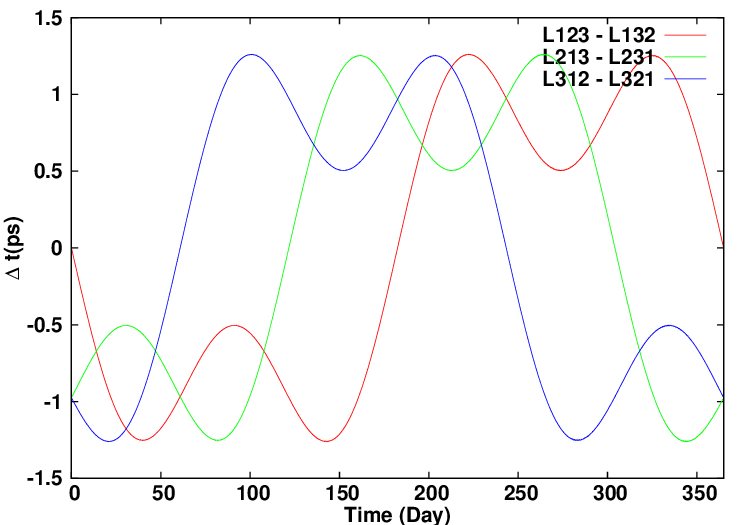}\includegraphics[width=0.45\textwidth]{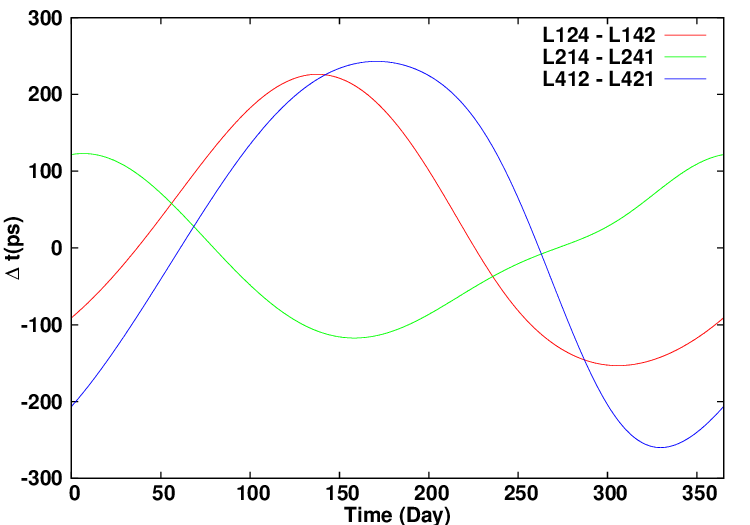} \includegraphics[width=0.45\textwidth]{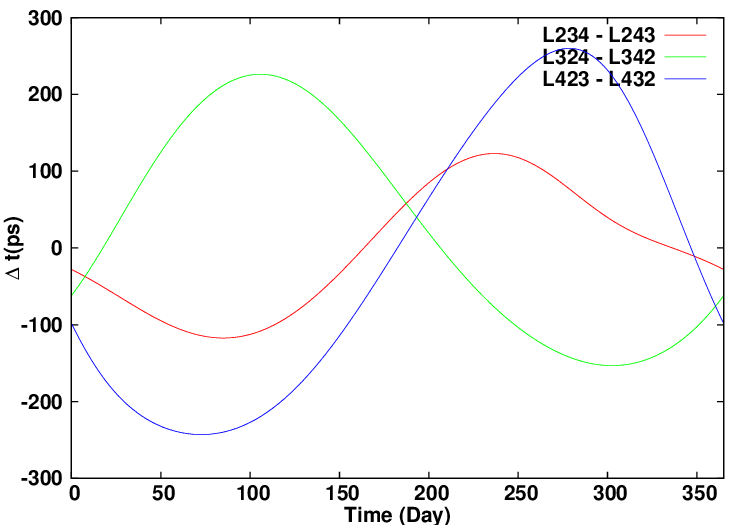}\includegraphics[width=0.45\textwidth]{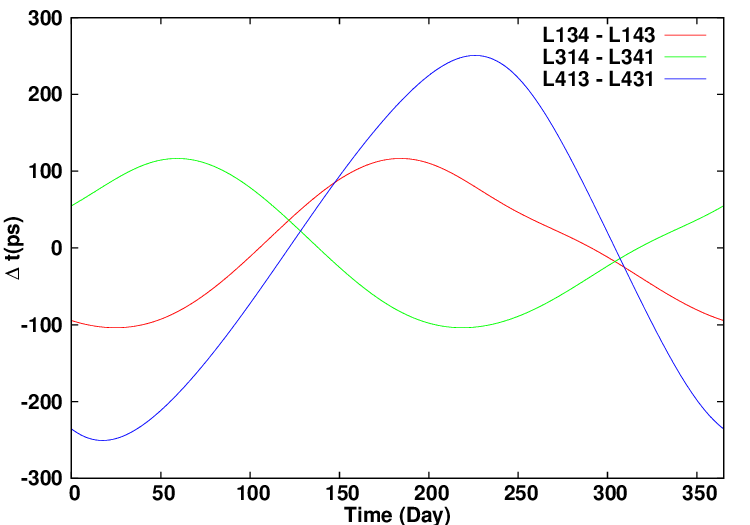}
    
    \caption{The optical path differences of the second-generation Sagnac TDI in one triangle of S/Cs over a one-year orbital period. The unit of the y-coordinate is the picosecond (ps) and the optical path differences is C$\Delta t$. C is light speed. As an example, in the legend, L123 - L132 means the optical path differences between S/C1 $\rightarrow$ S/C2 $\rightarrow$ S/C3 $\rightarrow$ S/C1 $\rightarrow$ S/C3 $\rightarrow$ S/C2 $\rightarrow$ S/C1 and S/C1 $\rightarrow$ S/C3 $\rightarrow$ S/C2 $\rightarrow$ S/C1 $\rightarrow$ S/C2 $\rightarrow$ S/C3 $\rightarrow$ S/C1.}
    \label{fig:Sagnac}
\end{figure*}

\begin{figure*}[!htbp]
	\centering 
     \includegraphics[width=0.45\textwidth]{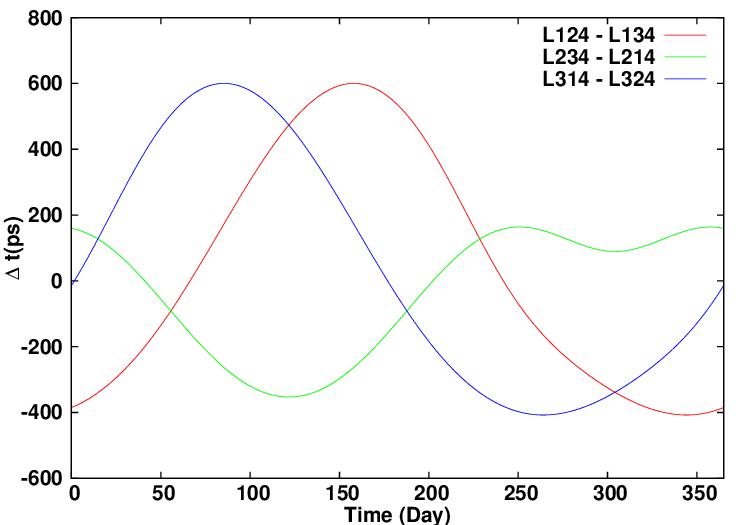}\includegraphics[width=0.45\textwidth]{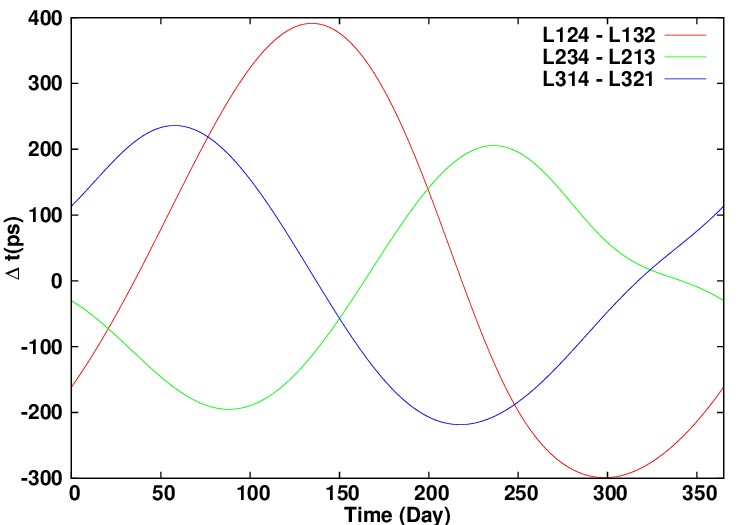}
    \caption{The optical path differences of the second-generation Sagnac TDI in two triangles of S/Cs over a one-year orbital period. The unit of the y-coordinate is the picosecond (ps) and the optical path differences is C$\Delta t$. C is light speed. As an example, in the legend, L124 - L134 means the optical path differences between S/C1 $\rightarrow$ S/C2 $\rightarrow$ S/C4 $\rightarrow$ S/C1 $\rightarrow$ S/C3 $\rightarrow$ S/C4 $\rightarrow$ S/C1 $\rightarrow$ S/C4 $\rightarrow$ S/C2  $\rightarrow$ S/C1 $\rightarrow$ S/C4 $\rightarrow$ S/C3 $\rightarrow$ S/C1 and S/C1 $\rightarrow$ S/C4 $\rightarrow$ S/C2  $\rightarrow$ S/C1 $\rightarrow$ S/C4 $\rightarrow$ S/C3 $\rightarrow$ S/C1$\rightarrow$ S/C2 $\rightarrow$ S/C4 $\rightarrow$ S/C1 $\rightarrow$ S/C3 $\rightarrow$ S/C4 $\rightarrow$ S/C1.}
    \label{fig:SagnacMix}
\end{figure*}

In the Sagnac TDI configuration for three S/Cs, the optical path differences of the second-generation TDI are shown in Figure \ref{fig:Sagnac} and \ref{fig:SagnacMix}. In one triangle of S/C1, S/C2, and S/C3, the optical path differences are roughly $1/10$ less than Michelson TDI and the other optical path differences are consistent with the ones relevant to Michelson TDI, which is shown in Figure \ref{fig:Sagnac}. That is derived from the variable arm lengths relevant to S/C4. When the combinations of Sagnac TDI are in two triangles of S/Cs, the optical path differences of the double combinations are almost close to the one triangles of S/Cs.The results and combinations are found in Figure \ref{fig:SagnacMix}.

\begin{figure*}[!htbp]
	\centering 
    \includegraphics[width=0.3\textwidth]{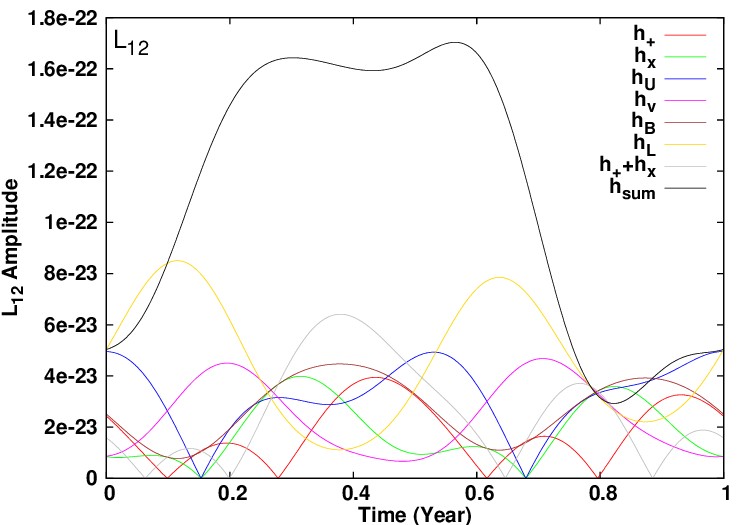}
    \includegraphics[width=0.3\textwidth]{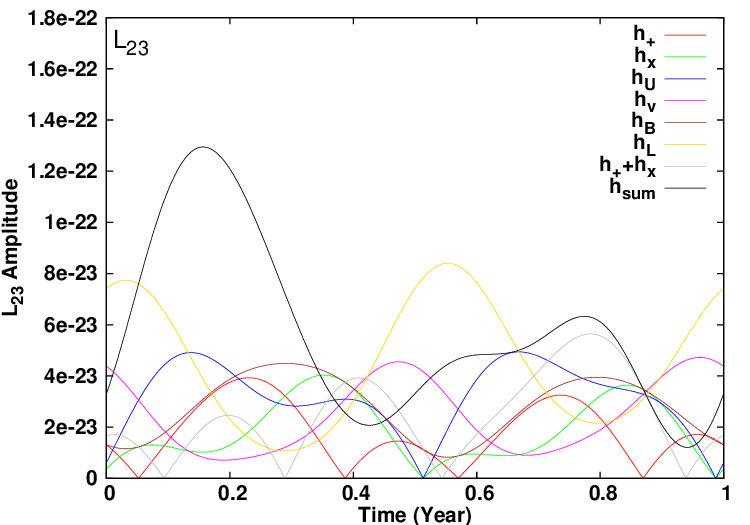}
    \includegraphics[width=0.3\textwidth]{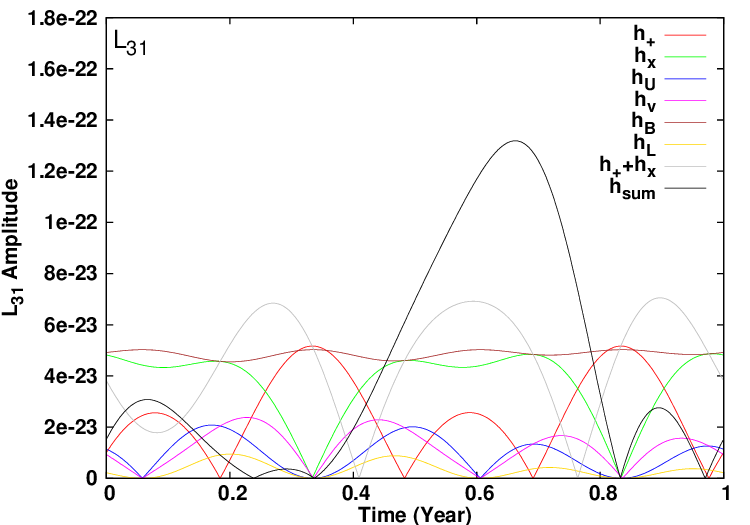}
    \includegraphics[width=0.3\textwidth]{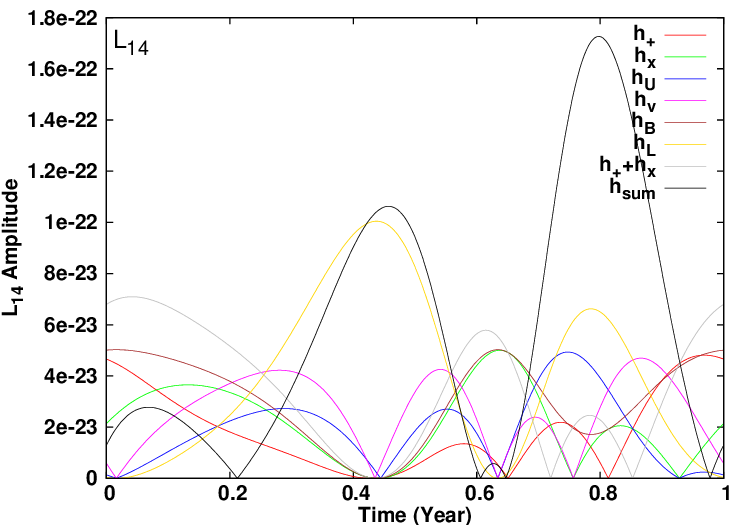}
    \includegraphics[width=0.3\textwidth]{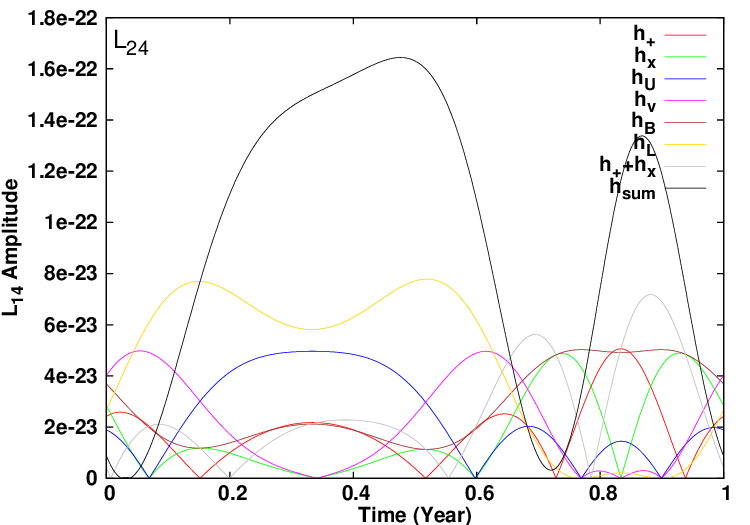}
    \includegraphics[width=0.3\textwidth]{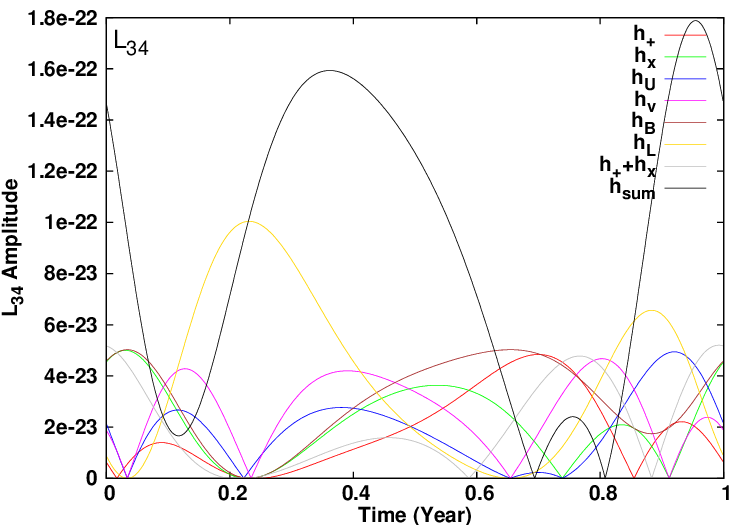}
    \caption{Amplitude of six polarization modes of GW source: J0806 response to all the arms of TEGO over a one-year orbital period. As an example, $L_{12}$ means the arm between S/C1 and S/C2.}
    \label{fig:polarizations}
\end{figure*}

In summary, as the radial distance of S/C4 is longer than the other S/Cs, three arm lengths relevant to S/C4 are changed apparently over a one-year orbital period. The second-generation TDI are selected to eliminate the laser frequency noise, instead of the first generation TDI. Comparing to the optical path differences  shown in Figure \ref{fig:Michelson}, \ref{fig:Sagnac} and \ref{fig:SagnacMix}, the unequal arm Michelson TDI configuration and the Sagnac TDI configuration for three S/Cs are equally effective for elimination of the laser frequency noise. 

A verification white dwarf binary signal: J0806 as the identified source is selected, whose parameters are detailed in the paper\citep{Kupfer:2018jee}. The polarization angle is set $\psi=0$, and initial orbital phase is set $\phi_0=0$. The sky location parameters \mbox{$\phi [rad], \beta [rad]$ and $ D_L [kpc]$} are 2.1021, -0.0821 and 5 respectively\citep{Kupfer:2018jee}. As a monochromatic source, The GW waveform of J0806 is calculated using the quadrupole approximation \citep{Landau:1962,Peters:1963}. Based on this approximation, the GW signals are described as a combination of the two polarizations ($+, \times$)\citep{Korol:2021pun} and the other four components of GW polarizations cannot be calculated with that approximation. In this paper, the GW waveforms relevant to the other four components of GW polarizations are set to be equal to ($\times$) component, artificially. That is an independent model.  The details are found in the subsection of Method section: GW Signal Model \ref{wavformModel}. The every components of GW polarization expression in the subsection of Method section: Polarization Modes \ref{polarizationModel}. 

As seen in Figure~\ref{fig:polarizations}, all the detector arms are simultaneously sensitive to the six polarization modes of GWs. As the angle between a detector arm and the orientation of a polarization mode changes with the detector's orbital position, a polarization mode of GWs exhibits variable amplitude over a one-year orbital period. This implies that there is an optimal orbital position for the detector where the amplitude of a given polarization mode of GWs is maximized.  For instance, a scalar longitudinal mode of GWs beyond GR has been identified as a dominant polarization component from the arm response between S/C1 and S/C4, and S/C3 and S/C4, at certain orbital positions.
The GW responses between S/C4 and the arms S/C1, S/C2, and S/C3 differ from each other. As an observable quantity, $h_{\text{sum}}$ represents the amplitude of the GW response in a detector arm. $h_{\text{sum}}$ varies with the detector's orbital position over a one-year period. The three newly added arms provide a distinct GW response compared to the traditional triangular configuration, as observed in projects such as LISA, Taiji, and Tianqin. This distinction offers additional degrees of freedom for extracting the polarization components from the total GW signals.
\section{Conclusion} 
In conclusion, the Tetrahedron Constellation of Gravitational Wave Observatory (TEGO) is a novel system comprising four spacecrafts (S/Cs) that enhances sensitivity to gravitational wave polarizations beyond traditional ground-based or space observatories like LIGO, Virgo, KAGRA, LISA, Taiji, and TianQin.
The laser telescopes and their pointing structures are mounted on the S/C platform and are evenly distributed at three locations 120 degrees apart. Comparing to a three-S/C configuration, the addition of a fourth S/C and its mounted telescope serve as redundant backups, especially for the telescopes, which together with the other two telescopes, automatically form a stable mass center for the platform without the need for trim mass.
The TEGO employs more combinations of time delay interferometry configurations to effectively suppress laser frequency noise, offering more sensitivity to gravitational wave signals.
As the configuration of TEGO is simultaneously sensitive to the six polarization modes of GWs, the TEGO allows for greater flexibility in extracting polarization components from total GW signals, making it simultaneously sensitive to all six polarization modes of gravitational waves.
%\end{methods}
%
%\subsection{Data Availability}
%
%\subsection{Code Availability}

%\begin{addendum}
\section*{Acknowledgements}
We thank the anonymous referees for very helpful suggestions, which are used to improve the manuscript.
We thank Yue-Liang Wu for the theory of GW polarization beyond GR and requirement of the space mission of GW detection, which are useful to improve the manuscript.  
This work has been supported by the Fundamental Research Funds for the Central Universities.
This work has been supported in part by the Strategic Priority Research Program of the Chinese Academy of Sciences under Grant No. XDA15020701 and XDA15020708. In this paper, the numerical computation of this work was completed at TAIJI Cluster of University of Chinese Academy of Sciences. 
\section*{Author contributions}
   Hong-Bo Jin contributed to project designing, data preparation, analysis and manuscript writing. 
   
   Cong-Feng Qiao contributed to project  designing, data preparation, analysis and manuscript writing. 
   
   All authors reviewed and commented on the final manuscript.
\section*{Competing Interests} The authors declare no competing interests.
%\item[Correspondence] Correspondence and requests for materials
% should be addressed to Cong-Feng Qiao(qiaocf@ucas.ac.cn) 

%\end{addendum}
%\clearpage
%\section{References}
\bibliographystyle{apsrev}
\bibliography{library,ref}

\begin{thebibliography}{30}
\expandafter\ifx\csname natexlab\endcsname\relax\def\natexlab#1{#1}\fi
\expandafter\ifx\csname bibnamefont\endcsname\relax
  \def\bibnamefont#1{#1}\fi
\expandafter\ifx\csname bibfnamefont\endcsname\relax
  \def\bibfnamefont#1{#1}\fi
\expandafter\ifx\csname citenamefont\endcsname\relax
  \def\citenamefont#1{#1}\fi
\expandafter\ifx\csname url\endcsname\relax
  \def\url#1{\texttt{#1}}\fi
\expandafter\ifx\csname urlprefix\endcsname\relax\def\urlprefix{URL }\fi
\providecommand{\bibinfo}[2]{#2}
\providecommand{\eprint}[2][]{\url{#2}}

\bibitem[{\citenamefont{Stevenson et~al.}(2017)\citenamefont{Stevenson,
  Vigna-G{\'{o}}mez, Mandel, Barrett, Neijssel, Perkins, and
  de~Mink}}]{2017NatCo...814906S}
\bibinfo{author}{\bibfnamefont{S.}~\bibnamefont{Stevenson}},
  \bibinfo{author}{\bibfnamefont{A.}~\bibnamefont{Vigna-G{\'{o}}mez}},
  \bibinfo{author}{\bibfnamefont{I.}~\bibnamefont{Mandel}},
  \bibinfo{author}{\bibfnamefont{J.~W.} \bibnamefont{Barrett}},
  \bibinfo{author}{\bibfnamefont{C.~J.} \bibnamefont{Neijssel}},
  \bibinfo{author}{\bibfnamefont{D.}~\bibnamefont{Perkins}}, \bibnamefont{and}
  \bibinfo{author}{\bibfnamefont{S.~E.} \bibnamefont{de~Mink}},
  \bibinfo{journal}{Nature Communications} \textbf{\bibinfo{volume}{8}},
  \bibinfo{pages}{14906} (\bibinfo{year}{2017}), ISSN
  \bibinfo{issn}{2041-1723}, \eprint{1704.01352},
  \urlprefix\url{http://www.nature.com/articles/ncomms14906}.

\bibitem[{\citenamefont{Acernese et~al.}(2019)\citenamefont{Acernese, Agathos,
  Aiello, Allocca, Amato, Ansoldi, Antier, Ar{\`{e}}ne, Arnaud, Ascenzi
  et~al.}}]{Virgo:2019juy}
\bibinfo{author}{\bibfnamefont{F.}~\bibnamefont{Acernese}},
  \bibinfo{author}{\bibfnamefont{M.}~\bibnamefont{Agathos}},
  \bibinfo{author}{\bibfnamefont{L.}~\bibnamefont{Aiello}},
  \bibinfo{author}{\bibfnamefont{A.}~\bibnamefont{Allocca}},
  \bibinfo{author}{\bibfnamefont{A.}~\bibnamefont{Amato}},
  \bibinfo{author}{\bibfnamefont{S.}~\bibnamefont{Ansoldi}},
  \bibinfo{author}{\bibfnamefont{S.}~\bibnamefont{Antier}},
  \bibinfo{author}{\bibfnamefont{M.}~\bibnamefont{Ar{\`{e}}ne}},
  \bibinfo{author}{\bibfnamefont{N.}~\bibnamefont{Arnaud}},
  \bibinfo{author}{\bibfnamefont{S.}~\bibnamefont{Ascenzi}},
  \bibnamefont{et~al.}, \bibinfo{journal}{Physical Review Letters}
  \textbf{\bibinfo{volume}{123}}, \bibinfo{pages}{231108}
  (\bibinfo{year}{2019}), ISSN \bibinfo{issn}{0031-9007},
  \urlprefix\url{https://link.aps.org/doi/10.1103/PhysRevLett.123.231108}.

\bibitem[{\citenamefont{Akutsu et~al.}(2019)\citenamefont{Akutsu, Ando, Arai,
  Arai, Araki, Araya, Aritomi, Asada, Aso, Atsuta et~al.}}]{KAGRA:2018plz}
\bibinfo{author}{\bibfnamefont{T.}~\bibnamefont{Akutsu}},
  \bibinfo{author}{\bibfnamefont{M.}~\bibnamefont{Ando}},
  \bibinfo{author}{\bibfnamefont{K.}~\bibnamefont{Arai}},
  \bibinfo{author}{\bibfnamefont{Y.}~\bibnamefont{Arai}},
  \bibinfo{author}{\bibfnamefont{S.}~\bibnamefont{Araki}},
  \bibinfo{author}{\bibfnamefont{A.}~\bibnamefont{Araya}},
  \bibinfo{author}{\bibfnamefont{N.}~\bibnamefont{Aritomi}},
  \bibinfo{author}{\bibfnamefont{H.}~\bibnamefont{Asada}},
  \bibinfo{author}{\bibfnamefont{Y.}~\bibnamefont{Aso}},
  \bibinfo{author}{\bibfnamefont{S.}~\bibnamefont{Atsuta}},
  \bibnamefont{et~al.}, \bibinfo{journal}{Nature Astronomy}
  \textbf{\bibinfo{volume}{3}}, \bibinfo{pages}{35} (\bibinfo{year}{2019}),
  ISSN \bibinfo{issn}{2397-3366}, \eprint{1811.08079},
  \urlprefix\url{https://www.nature.com/articles/s41550-018-0658-y}.

\bibitem[{\citenamefont{Hagihara et~al.}(2018)\citenamefont{Hagihara, Era,
  Iikawa, and Asada}}]{Hagihara:2018azu}
\bibinfo{author}{\bibfnamefont{Y.}~\bibnamefont{Hagihara}},
  \bibinfo{author}{\bibfnamefont{N.}~\bibnamefont{Era}},
  \bibinfo{author}{\bibfnamefont{D.}~\bibnamefont{Iikawa}}, \bibnamefont{and}
  \bibinfo{author}{\bibfnamefont{H.}~\bibnamefont{Asada}},
  \bibinfo{journal}{Physical Review D} \textbf{\bibinfo{volume}{98}},
  \bibinfo{pages}{064035} (\bibinfo{year}{2018}), ISSN
  \bibinfo{issn}{2470-0010}, \eprint{1807.07234},
  \urlprefix\url{https://link.aps.org/doi/10.1103/PhysRevD.98.064035}.

\bibitem[{\citenamefont{Hagihara et~al.}(2019)\citenamefont{Hagihara, Era,
  Iikawa, Nishizawa, and Asada}}]{Hagihara:2019ihn}
\bibinfo{author}{\bibfnamefont{Y.}~\bibnamefont{Hagihara}},
  \bibinfo{author}{\bibfnamefont{N.}~\bibnamefont{Era}},
  \bibinfo{author}{\bibfnamefont{D.}~\bibnamefont{Iikawa}},
  \bibinfo{author}{\bibfnamefont{A.}~\bibnamefont{Nishizawa}},
  \bibnamefont{and} \bibinfo{author}{\bibfnamefont{H.}~\bibnamefont{Asada}},
  \bibinfo{journal}{Physical Review D} \textbf{\bibinfo{volume}{100}},
  \bibinfo{pages}{064010} (\bibinfo{year}{2019}), ISSN
  \bibinfo{issn}{2470-0010}, \eprint{1904.02300},
  \urlprefix\url{https://link.aps.org/doi/10.1103/PhysRevD.100.064010}.

\bibitem[{\citenamefont{Collaboration}(2021)}]{TaijiScientific:2021qgx}
\bibinfo{author}{\bibfnamefont{T.~T.~S.} \bibnamefont{Collaboration}},
  \bibinfo{journal}{Communications Physics} \textbf{\bibinfo{volume}{4}},
  \bibinfo{pages}{34} (\bibinfo{year}{2021}), ISSN \bibinfo{issn}{2399-3650},
  \urlprefix\url{http://www.nature.com/articles/s42005-021-00529-z}.

\bibitem[{\citenamefont{Gong et~al.}(2021)\citenamefont{Gong, Luo, and
  Wang}}]{Gong:2021gvw}
\bibinfo{author}{\bibfnamefont{Y.}~\bibnamefont{Gong}},
  \bibinfo{author}{\bibfnamefont{J.}~\bibnamefont{Luo}}, \bibnamefont{and}
  \bibinfo{author}{\bibfnamefont{B.}~\bibnamefont{Wang}},
  \bibinfo{journal}{Nature Astronomy} \textbf{\bibinfo{volume}{5}},
  \bibinfo{pages}{881} (\bibinfo{year}{2021}), ISSN \bibinfo{issn}{2397-3366},
  \eprint{2109.07442},
  \urlprefix\url{https://www.nature.com/articles/s41550-021-01480-3}.

\bibitem[{\citenamefont{Liu et~al.}(2020)\citenamefont{Liu, Ruan, and
  Guo}}]{Liu:2020mab}
\bibinfo{author}{\bibfnamefont{C.}~\bibnamefont{Liu}},
  \bibinfo{author}{\bibfnamefont{W.-H.} \bibnamefont{Ruan}}, \bibnamefont{and}
  \bibinfo{author}{\bibfnamefont{Z.-K.} \bibnamefont{Guo}},
  \bibinfo{journal}{Physical Review D} \textbf{\bibinfo{volume}{102}},
  \bibinfo{pages}{124050} (\bibinfo{year}{2020}), ISSN
  \bibinfo{issn}{2470-0010}, \eprint{2006.04413},
  \urlprefix\url{http://dx.doi.org/10.1103/PhysRevD.102.124050}.

\bibitem[{\citenamefont{Hu et~al.}(2024)\citenamefont{Hu, Wang, Tan, and
  Shao}}]{Hu:2024toa}
\bibinfo{author}{\bibfnamefont{Y.}~\bibnamefont{Hu}},
  \bibinfo{author}{\bibfnamefont{P.-P.} \bibnamefont{Wang}},
  \bibinfo{author}{\bibfnamefont{Y.-J.} \bibnamefont{Tan}}, \bibnamefont{and}
  \bibinfo{author}{\bibfnamefont{C.-G.} \bibnamefont{Shao}},
  \bibinfo{journal}{The Astrophysical Journal} \textbf{\bibinfo{volume}{961}},
  \bibinfo{pages}{116} (\bibinfo{year}{2024}), ISSN \bibinfo{issn}{0004-637X},
  \urlprefix\url{https://iopscience.iop.org/article/10.3847/1538-4357/ad0cef}.

\bibitem[{\citenamefont{Nishizawa et~al.}(2009)\citenamefont{Nishizawa, Taruya,
  Hayama, Kawamura, and Sakagami}}]{Nishizawa:2009bf}
\bibinfo{author}{\bibfnamefont{A.}~\bibnamefont{Nishizawa}},
  \bibinfo{author}{\bibfnamefont{A.}~\bibnamefont{Taruya}},
  \bibinfo{author}{\bibfnamefont{K.}~\bibnamefont{Hayama}},
  \bibinfo{author}{\bibfnamefont{S.}~\bibnamefont{Kawamura}}, \bibnamefont{and}
  \bibinfo{author}{\bibfnamefont{M.-a.} \bibnamefont{Sakagami}},
  \bibinfo{journal}{Physical Review D} \textbf{\bibinfo{volume}{79}},
  \bibinfo{pages}{082002} (\bibinfo{year}{2009}), ISSN
  \bibinfo{issn}{1550-7998}, \eprint{0903.0528},
  \urlprefix\url{https://link.aps.org/doi/10.1103/PhysRevD.79.082002}.

\bibitem[{\citenamefont{Wang et~al.}(2013)\citenamefont{Wang, Keitel, Babak,
  Petiteau, Otto, Barke, Kawazoe, Khalaidovski, M{\"{u}}ller, Sch{\"{u}}tze
  et~al.}}]{Wang:2013lna}
\bibinfo{author}{\bibfnamefont{Y.}~\bibnamefont{Wang}},
  \bibinfo{author}{\bibfnamefont{D.}~\bibnamefont{Keitel}},
  \bibinfo{author}{\bibfnamefont{S.}~\bibnamefont{Babak}},
  \bibinfo{author}{\bibfnamefont{A.}~\bibnamefont{Petiteau}},
  \bibinfo{author}{\bibfnamefont{M.}~\bibnamefont{Otto}},
  \bibinfo{author}{\bibfnamefont{S.}~\bibnamefont{Barke}},
  \bibinfo{author}{\bibfnamefont{F.}~\bibnamefont{Kawazoe}},
  \bibinfo{author}{\bibfnamefont{A.}~\bibnamefont{Khalaidovski}},
  \bibinfo{author}{\bibfnamefont{V.}~\bibnamefont{M{\"{u}}ller}},
  \bibinfo{author}{\bibfnamefont{D.}~\bibnamefont{Sch{\"{u}}tze}},
  \bibnamefont{et~al.}, \bibinfo{journal}{Physical Review D}
  \textbf{\bibinfo{volume}{88}}, \bibinfo{pages}{104021}
  (\bibinfo{year}{2013}), ISSN \bibinfo{issn}{1550-7998}, \eprint{1306.3865},
  \urlprefix\url{https://link.aps.org/doi/10.1103/PhysRevD.88.104021}.

\bibitem[{\citenamefont{Amaro-Seoane et~al.}(2017)\citenamefont{Amaro-Seoane,
  Audley, Babak, Baker, Barausse, Bender, Berti, Binetruy, Born, Bortoluzzi
  et~al.}}]{LISA:2017pwj}
\bibinfo{author}{\bibfnamefont{P.}~\bibnamefont{Amaro-Seoane}},
  \bibinfo{author}{\bibfnamefont{H.}~\bibnamefont{Audley}},
  \bibinfo{author}{\bibfnamefont{S.}~\bibnamefont{Babak}},
  \bibinfo{author}{\bibfnamefont{J.}~\bibnamefont{Baker}},
  \bibinfo{author}{\bibfnamefont{E.}~\bibnamefont{Barausse}},
  \bibinfo{author}{\bibfnamefont{P.}~\bibnamefont{Bender}},
  \bibinfo{author}{\bibfnamefont{E.}~\bibnamefont{Berti}},
  \bibinfo{author}{\bibfnamefont{P.}~\bibnamefont{Binetruy}},
  \bibinfo{author}{\bibfnamefont{M.}~\bibnamefont{Born}},
  \bibinfo{author}{\bibfnamefont{D.}~\bibnamefont{Bortoluzzi}},
  \bibnamefont{et~al.}, \bibinfo{journal}{arXiv preprint arXiv:1702.00786}
  (\bibinfo{year}{2017}).

\bibitem[{\citenamefont{Bayle et~al.}(2022)\citenamefont{Bayle, Bonga, Caprini,
  Doneva, Muratore, Petiteau, Rossi, and Shao}}]{2022NatAs...6.1334B}
\bibinfo{author}{\bibfnamefont{J.-B.} \bibnamefont{Bayle}},
  \bibinfo{author}{\bibfnamefont{B.}~\bibnamefont{Bonga}},
  \bibinfo{author}{\bibfnamefont{C.}~\bibnamefont{Caprini}},
  \bibinfo{author}{\bibfnamefont{D.}~\bibnamefont{Doneva}},
  \bibinfo{author}{\bibfnamefont{M.}~\bibnamefont{Muratore}},
  \bibinfo{author}{\bibfnamefont{A.}~\bibnamefont{Petiteau}},
  \bibinfo{author}{\bibfnamefont{E.}~\bibnamefont{Rossi}}, \bibnamefont{and}
  \bibinfo{author}{\bibfnamefont{L.}~\bibnamefont{Shao}},
  \bibinfo{journal}{Nature Astronomy} \textbf{\bibinfo{volume}{6}},
  \bibinfo{pages}{1334} (\bibinfo{year}{2022}), ISSN \bibinfo{issn}{2397-3366},
  \urlprefix\url{https://www.nature.com/articles/s41550-022-01847-0}.

\bibitem[{\citenamefont{Hu and Wu}(2017)}]{Hu:2017mde}
\bibinfo{author}{\bibfnamefont{W.-R.} \bibnamefont{Hu}} \bibnamefont{and}
  \bibinfo{author}{\bibfnamefont{Y.-L.} \bibnamefont{Wu}},
  \bibinfo{journal}{National Science Review} \textbf{\bibinfo{volume}{4}},
  \bibinfo{pages}{685} (\bibinfo{year}{2017}), ISSN \bibinfo{issn}{2095-5138},
  \urlprefix\url{https://academic.oup.com/nsr/article/4/5/685/4430188}.

\bibitem[{\citenamefont{Ruan et~al.}(2020)\citenamefont{Ruan, Liu, Guo, Wu, and
  Cai}}]{Ruan:2020smc}
\bibinfo{author}{\bibfnamefont{W.-H.} \bibnamefont{Ruan}},
  \bibinfo{author}{\bibfnamefont{C.}~\bibnamefont{Liu}},
  \bibinfo{author}{\bibfnamefont{Z.-K.} \bibnamefont{Guo}},
  \bibinfo{author}{\bibfnamefont{Y.-L.} \bibnamefont{Wu}}, \bibnamefont{and}
  \bibinfo{author}{\bibfnamefont{R.-G.} \bibnamefont{Cai}},
  \bibinfo{journal}{Nature Astronomy} \textbf{\bibinfo{volume}{4}},
  \bibinfo{pages}{108} (\bibinfo{year}{2020}), ISSN \bibinfo{issn}{2397-3366},
  \eprint{2002.03603},
  \urlprefix\url{http://www.nature.com/articles/s41550-019-1008-4}.

\bibitem[{\citenamefont{Luo et~al.}(2015)\citenamefont{Luo, Chen, Duan, Gong,
  Hu, Ji, Liu, Mei, Milyukov, Sazhin et~al.}}]{TianQin:2015yph}
\bibinfo{author}{\bibfnamefont{J.}~\bibnamefont{Luo}},
  \bibinfo{author}{\bibfnamefont{L.-S.} \bibnamefont{Chen}},
  \bibinfo{author}{\bibfnamefont{H.-Z.} \bibnamefont{Duan}},
  \bibinfo{author}{\bibfnamefont{Y.-G.} \bibnamefont{Gong}},
  \bibinfo{author}{\bibfnamefont{S.}~\bibnamefont{Hu}},
  \bibinfo{author}{\bibfnamefont{J.}~\bibnamefont{Ji}},
  \bibinfo{author}{\bibfnamefont{Q.}~\bibnamefont{Liu}},
  \bibinfo{author}{\bibfnamefont{J.}~\bibnamefont{Mei}},
  \bibinfo{author}{\bibfnamefont{V.}~\bibnamefont{Milyukov}},
  \bibinfo{author}{\bibfnamefont{M.}~\bibnamefont{Sazhin}},
  \bibnamefont{et~al.}, \bibinfo{journal}{Classical and Quantum Gravity}
  \textbf{\bibinfo{volume}{33}}, \bibinfo{pages}{035010}
  (\bibinfo{year}{2015}), ISSN \bibinfo{issn}{0264-9381}, \eprint{1512.02076},
  \urlprefix\url{http://dx.doi.org/10.1088/0264-9381/33/3/035010}.

\bibitem[{\citenamefont{Mei et~al.}(2020)\citenamefont{Mei, Bai, Bao, Barausse,
  Cai, Canuto, Cao, Chen, Chen, Ding et~al.}}]{TianQin:2020hid}
\bibinfo{author}{\bibfnamefont{J.}~\bibnamefont{Mei}},
  \bibinfo{author}{\bibfnamefont{Y.-Z.} \bibnamefont{Bai}},
  \bibinfo{author}{\bibfnamefont{J.}~\bibnamefont{Bao}},
  \bibinfo{author}{\bibfnamefont{E.}~\bibnamefont{Barausse}},
  \bibinfo{author}{\bibfnamefont{L.}~\bibnamefont{Cai}},
  \bibinfo{author}{\bibfnamefont{E.}~\bibnamefont{Canuto}},
  \bibinfo{author}{\bibfnamefont{B.}~\bibnamefont{Cao}},
  \bibinfo{author}{\bibfnamefont{W.-M.} \bibnamefont{Chen}},
  \bibinfo{author}{\bibfnamefont{Y.}~\bibnamefont{Chen}},
  \bibinfo{author}{\bibfnamefont{Y.-W.} \bibnamefont{Ding}},
  \bibnamefont{et~al.}, \bibinfo{journal}{Progress of Theoretical and
  Experimental Physics} \textbf{\bibinfo{volume}{2021}},
  \bibinfo{pages}{05A107} (\bibinfo{year}{2020}), ISSN
  \bibinfo{issn}{2050-3911}, \eprint{2008.10332},
  \urlprefix\url{http://dx.doi.org/10.1093/ptep/ptaa114}.

\bibitem[{\citenamefont{Dhurandhar et~al.}(2013)\citenamefont{Dhurandhar, Ni,
  and Wang}}]{Dhurandhar:2011ik}
\bibinfo{author}{\bibfnamefont{S.}~\bibnamefont{Dhurandhar}},
  \bibinfo{author}{\bibfnamefont{W.-T.} \bibnamefont{Ni}}, \bibnamefont{and}
  \bibinfo{author}{\bibfnamefont{G.}~\bibnamefont{Wang}},
  \bibinfo{journal}{Advances in Space Research} \textbf{\bibinfo{volume}{51}},
  \bibinfo{pages}{198} (\bibinfo{year}{2013}), ISSN \bibinfo{issn}{02731177},
  \eprint{1102.4965},
  \urlprefix\url{https://linkinghub.elsevier.com/retrieve/pii/S0273117712005893}.

\bibitem[{\citenamefont{Tinto and Dhurandhar}(2014)}]{Tinto:2014lxa}
\bibinfo{author}{\bibfnamefont{M.}~\bibnamefont{Tinto}} \bibnamefont{and}
  \bibinfo{author}{\bibfnamefont{S.~V.} \bibnamefont{Dhurandhar}},
  \bibinfo{journal}{Living Reviews in Relativity}
  \textbf{\bibinfo{volume}{17}}, \bibinfo{pages}{6} (\bibinfo{year}{2014}),
  ISSN \bibinfo{issn}{2367-3613},
  \urlprefix\url{https://link.springer.com/10.12942/lrr-2014-6}.

\bibitem[{\citenamefont{Wang and Ni}(2019)}]{Wang:2017aqq}
\bibinfo{author}{\bibfnamefont{G.}~\bibnamefont{Wang}} \bibnamefont{and}
  \bibinfo{author}{\bibfnamefont{W.-T.} \bibnamefont{Ni}},
  \bibinfo{journal}{Research in Astronomy and Astrophysics}
  \textbf{\bibinfo{volume}{19}}, \bibinfo{pages}{058} (\bibinfo{year}{2019}),
  ISSN \bibinfo{issn}{1674-4527}, \eprint{1707.09127},
  \urlprefix\url{http://dx.doi.org/10.1088/1674-4527/19/4/58}.

\bibitem[{\citenamefont{Vallisneri}(2005)}]{Vallisneri:2005ji}
\bibinfo{author}{\bibfnamefont{M.}~\bibnamefont{Vallisneri}},
  \bibinfo{journal}{Physical Review D} \textbf{\bibinfo{volume}{76}},
  \bibinfo{pages}{109903} (\bibinfo{year}{2005}), ISSN
  \bibinfo{issn}{1550-7998}, \eprint{0504145},
  \urlprefix\url{https://link.aps.org/doi/10.1103/PhysRevD.76.109903}.

\bibitem[{\citenamefont{Rubbo et~al.}(2004)\citenamefont{Rubbo, Cornish, and
  Poujade}}]{Rubbo:2003ap}
\bibinfo{author}{\bibfnamefont{L.~J.} \bibnamefont{Rubbo}},
  \bibinfo{author}{\bibfnamefont{N.~J.} \bibnamefont{Cornish}},
  \bibnamefont{and} \bibinfo{author}{\bibfnamefont{O.}~\bibnamefont{Poujade}},
  \bibinfo{journal}{Physical Review D} \textbf{\bibinfo{volume}{69}},
  \bibinfo{pages}{082003} (\bibinfo{year}{2004}), ISSN
  \bibinfo{issn}{1550-7998}, \eprint{0311069},
  \urlprefix\url{https://link.aps.org/doi/10.1103/PhysRevD.69.082003}.

\bibitem[{\citenamefont{Cutler}(1998)}]{Cutler:1997ta}
\bibinfo{author}{\bibfnamefont{C.}~\bibnamefont{Cutler}},
  \bibinfo{journal}{Physical Review D} \textbf{\bibinfo{volume}{57}},
  \bibinfo{pages}{7089} (\bibinfo{year}{1998}), ISSN \bibinfo{issn}{0556-2821},
  \eprint{gr-qc/9703068},
  \urlprefix\url{https://link.aps.org/doi/10.1103/PhysRevD.57.7089}.

\bibitem[{\citenamefont{Roebber et~al.}(2020)\citenamefont{Roebber, Buscicchio,
  Vecchio, Moore, Klein, Korol, Toonen, Gerosa, Goldstein, Gaebel
  et~al.}}]{roe20}
\bibinfo{author}{\bibfnamefont{E.}~\bibnamefont{Roebber}},
  \bibinfo{author}{\bibfnamefont{R.}~\bibnamefont{Buscicchio}},
  \bibinfo{author}{\bibfnamefont{A.}~\bibnamefont{Vecchio}},
  \bibinfo{author}{\bibfnamefont{C.~J.} \bibnamefont{Moore}},
  \bibinfo{author}{\bibfnamefont{A.}~\bibnamefont{Klein}},
  \bibinfo{author}{\bibfnamefont{V.}~\bibnamefont{Korol}},
  \bibinfo{author}{\bibfnamefont{S.}~\bibnamefont{Toonen}},
  \bibinfo{author}{\bibfnamefont{D.}~\bibnamefont{Gerosa}},
  \bibinfo{author}{\bibfnamefont{J.}~\bibnamefont{Goldstein}},
  \bibinfo{author}{\bibfnamefont{S.~M.} \bibnamefont{Gaebel}},
  \bibnamefont{et~al.}, \bibinfo{journal}{The Astrophysical Journal}
  \textbf{\bibinfo{volume}{894}}, \bibinfo{pages}{L15} (\bibinfo{year}{2020}),
  ISSN \bibinfo{issn}{2041-8213}, \eprint{2002.10465},
  \urlprefix\url{https://iopscience.iop.org/article/10.3847/2041-8213/ab8ac9}.

\bibitem[{\citenamefont{Karnesis et~al.}(2021)\citenamefont{Karnesis, Babak,
  Pieroni, Cornish, and Littenberg}}]{kar21}
\bibinfo{author}{\bibfnamefont{N.}~\bibnamefont{Karnesis}},
  \bibinfo{author}{\bibfnamefont{S.}~\bibnamefont{Babak}},
  \bibinfo{author}{\bibfnamefont{M.}~\bibnamefont{Pieroni}},
  \bibinfo{author}{\bibfnamefont{N.}~\bibnamefont{Cornish}}, \bibnamefont{and}
  \bibinfo{author}{\bibfnamefont{T.}~\bibnamefont{Littenberg}},
  \bibinfo{journal}{Physical Review D} \textbf{\bibinfo{volume}{104}},
  \bibinfo{pages}{043019} (\bibinfo{year}{2021}), ISSN
  \bibinfo{issn}{2470-0010},
  \urlprefix\url{https://link.aps.org/doi/10.1103/PhysRevD.104.043019}.

\bibitem[{\citenamefont{Landau and Lifshitz}(1962)}]{Landau:1962}
\bibinfo{author}{\bibfnamefont{L.~D.} \bibnamefont{Landau}} \bibnamefont{and}
  \bibinfo{author}{\bibfnamefont{E.~M.} \bibnamefont{Lifshitz}},
  \emph{\bibinfo{title}{{The classical theory of fields; 2nd ed.}}}, Course of
  theoretical physics (\bibinfo{publisher}{Pergamon},
  \bibinfo{address}{London}, \bibinfo{year}{1962}),
  \urlprefix\url{https://cds.cern.ch/record/101809}.

\bibitem[{\citenamefont{Peters and Mathews}(1963)}]{Peters:1963}
\bibinfo{author}{\bibfnamefont{P.~C.} \bibnamefont{Peters}} \bibnamefont{and}
  \bibinfo{author}{\bibfnamefont{J.}~\bibnamefont{Mathews}},
  \bibinfo{journal}{Phys. Rev.} \textbf{\bibinfo{volume}{131}},
  \bibinfo{pages}{435} (\bibinfo{year}{1963}),
  \urlprefix\url{https://link.aps.org/doi/10.1103/PhysRev.131.435}.

\bibitem[{\citenamefont{Korol et~al.}(2022)\citenamefont{Korol, Hallakoun,
  Toonen, and Karnesis}}]{Korol:2021pun}
\bibinfo{author}{\bibfnamefont{V.}~\bibnamefont{Korol}},
  \bibinfo{author}{\bibfnamefont{N.}~\bibnamefont{Hallakoun}},
  \bibinfo{author}{\bibfnamefont{S.}~\bibnamefont{Toonen}}, \bibnamefont{and}
  \bibinfo{author}{\bibfnamefont{N.}~\bibnamefont{Karnesis}},
  \bibinfo{journal}{Monthly Notices of the Royal Astronomical Society}
  \textbf{\bibinfo{volume}{511}}, \bibinfo{pages}{5936} (\bibinfo{year}{2022}),
  ISSN \bibinfo{issn}{0035-8711}, \eprint{2109.10972},
  \urlprefix\url{http://dx.doi.org/10.1093/mnras/stac415}.

\bibitem[{\citenamefont{Kupfer et~al.}(2018)\citenamefont{Kupfer, Korol, Shah,
  Nelemans, Marsh, Ramsay, Groot, Steeghs, and Rossi}}]{Kupfer:2018jee}
\bibinfo{author}{\bibfnamefont{T.}~\bibnamefont{Kupfer}},
  \bibinfo{author}{\bibfnamefont{V.}~\bibnamefont{Korol}},
  \bibinfo{author}{\bibfnamefont{S.}~\bibnamefont{Shah}},
  \bibinfo{author}{\bibfnamefont{G.}~\bibnamefont{Nelemans}},
  \bibinfo{author}{\bibfnamefont{T.~R.} \bibnamefont{Marsh}},
  \bibinfo{author}{\bibfnamefont{G.}~\bibnamefont{Ramsay}},
  \bibinfo{author}{\bibfnamefont{P.~J.} \bibnamefont{Groot}},
  \bibinfo{author}{\bibfnamefont{D.~T.~H.} \bibnamefont{Steeghs}},
  \bibnamefont{and} \bibinfo{author}{\bibfnamefont{E.~M.} \bibnamefont{Rossi}},
  \bibinfo{journal}{Monthly Notices of the Royal Astronomical Society}
  \textbf{\bibinfo{volume}{480}}, \bibinfo{pages}{302} (\bibinfo{year}{2018}),
  ISSN \bibinfo{issn}{0035-8711}, \eprint{1805.00482},
  \urlprefix\url{https://academic.oup.com/mnras/article/480/1/302/5037945}.

\bibitem[{\citenamefont{Guo et~al.}(2024)\citenamefont{Guo, Jin, Qiao, and
  Wu}}]{Guo:2023lzb}
\bibinfo{author}{\bibfnamefont{P.}~\bibnamefont{Guo}},
  \bibinfo{author}{\bibfnamefont{H.-B.} \bibnamefont{Jin}},
  \bibinfo{author}{\bibfnamefont{C.-F.} \bibnamefont{Qiao}}, \bibnamefont{and}
  \bibinfo{author}{\bibfnamefont{Y.-L.} \bibnamefont{Wu}},
  \bibinfo{journal}{Results in Physics} \textbf{\bibinfo{volume}{60}},
  \bibinfo{pages}{107607} (\bibinfo{year}{2024}), ISSN
  \bibinfo{issn}{22113797}, \eprint{2310.16796},
  \urlprefix\url{https://linkinghub.elsevier.com/retrieve/pii/S2211379724002900}.

\end{thebibliography}

\end{document}